\documentclass[
    ,final            
  ,mathptm                
  ]
  {aipproc}

\layoutstyle{8x11double}

\begin{document}

\title{Theory of Josephson Phenomena in Superfluid $^3$He}

\classification{67.57.De} \keywords{superfluid $^3$He, Josephson
effect, current--phase relation, $\pi$ state, Andreev bound state.}

\author{Erkki Thuneberg}{
  address={Department of Physical Sciences, 
University of Oulu, Finland} }

\begin{abstract}
Quite detailed theoretical description of superfluid  $^3$He is
possible on length scales that are much larger than the atomic scale.
We discuss weak links between two bulk states of $^3$He-B. The current
through the weak link is determined by the bound states at the link.
The bound state energies are spin split depending on the order
parameters in the bulk. As a result, unusual current--phase relations
with $\pi$ state appear. For not too weak links, the order
parameter in the bulk is modified because of the Josephson
coupling. This leads to a stronger $\pi$ state and to an additional
current at constant pressure bias. The theoretical results are 
compared with experiments.  
\end{abstract}

\maketitle

\section{Introduction}

In spite of more than thirty years of studies of superfluid phases of
$^3$He, there are several active research directions. One of them is
the study of Josephson phenomena in weak links between two volumes  of
bulk superfluid. Here we give a review of recent theoretical work on
this subject \cite{Viljas1,Viljas2,dynlett,Viljasoma}. We pay special
attention to the role of bound quasiparticle states in the link, and
explain how these determine the flow through the junction. In
particular we study the equilibrium current--phase relation, $I(\phi)$,
and the dc current at constant potential difference,
$I_{\rm dc}(U)$. We compare the results with experiments. The
experimental work has been reviewed in Ref.\
\cite{josrev}.  A theoretical review with a somewhat different emphasis
has been given in Ref.\
\cite{aitrev}.

\section{Weak Links}

Let us consider two volumes of superfluid that are connected by a weak
link. Here ``superfluid'' means generally a fermion superfluid, either
superconducting metal or liquid $^3$He. Our discussion of
superconducting weak links is mainly aimed as an introduction for
$^3$He. Therefore we neglect all complications that arise from impurity
or interfacial scattering in superconductors. Also we consider
only superconductors of the conventional (s-wave) type. In the case of
$^3$He we limit to the superfluid B phase. 

\begin{figure}\label{jostrajboundsc}
  \includegraphics[width=0.4\linewidth]{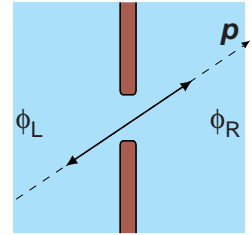}
  \caption{Sketch of a weak link between two bulk
  superfluids. The two bulk superfluids have phases $\phi_L$ and
  $\phi_R$. One quasiparticle trajectory with momentum
  ${\bf p}$ is shown. When its energy is within the energy gap, the
  quasiparticle cannot escape to the bulk but is Andreev reflected as
  hole with essentially the same momentum. This in turn is Andreev
  reflected as particle on the other side of the link. Thus particle and
  hole type excitations bounce back and forth along the same
       trajectory.  }
\end{figure}

The geometry of the weak link is
depicted in Fig.\ \ref{jostrajboundsc}.
The figure shows one quasiparticle trajectory with momentum ${\bf p}$.
In bulk superfluid such elementary excitations must have energy
$\epsilon$ larger than the energy gap $\Delta$ of the
superfluid. In the weak link, however, there exists energy  eigenvalues
within the energy gap, 
$\vert\epsilon\vert < \Delta$. In such a state the quasiparticle
cannot escape to the bulk but is Andreev reflected. In Andreev
reflection a particle type excitation is converted to hole type and
vice versa, with essentially unchanged momentum \cite{andreev}. Thus
the particles and holes travel the same trajectory back and forth,
respectively. In one cycle, one Cooper pair, or equivalently two
particles, are transmitted through the weak link.    

\begin{figure}\label{diffusedossc}
  \includegraphics[width=0.7\linewidth]{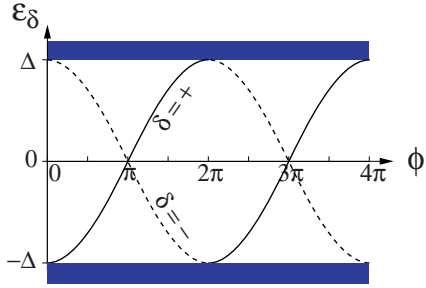}
  \caption{Bound state energies $\epsilon_\delta(\phi)$ in a
superconducting point contact. As $\phi$ increases, the states with
transport to the right ($\delta =+1$) have increasing
energy (solid lines), and the states with transport to the left ($\delta
=-1$) have decreasing energy (dashed lines). The states at energies
$\vert\epsilon\vert >\Delta$ form a continuum.}
\end{figure}

The bulk superfluid states are
characterized by phases
$\phi_L$ and
$\phi_R$. The energies of the bound states depend essentially on
the phase difference
$\phi=\phi_R-\phi_L$ \cite{Kulik}. For superconductors this is
depicted  in Fig.\ \ref{diffusedossc}.  
The states appear as pairs with positive and negative energies,
$\pm\vert\epsilon\vert$. Understanding that $d\epsilon/d\phi$ plays a
similar role as the group velocity, we can assign the states as
propagating to the  right ($\delta =+1$) and to the left ($\delta
=- 1$). These two states are not independent, however. Based on what was
stated about Andreev reflection above, if a state with $\delta=+1$ is
occupied, the corresponding $\delta=-1$ state must be empty. And vice
versa. More generally, the occupations $f_\pm$ have to satisfy
$f_++f_-=1$. Really the states with positive and negative energies are
the same bound state, whose  energy relative to the ground state is the
absolute value
$\vert\epsilon\vert$. Anyhow, we find it very useful to consider a
bound state as a superposition of positive and negative energy states.

In the simplest case of a point contact, the bound
state energies have the simple analytic expression 
$\epsilon=\pm\Delta\cos\frac{\phi}{2}$  \cite{Kulik}.  By a point
contact, or a pinhole in case of $^3$He, we mean a weak link whose all
dimensions are small compared to the superfluid coherence length
$\xi_0$.

It is an interesting observation that to a large extent the current
through a weak link is determined by the bound states
\cite{FT,Beenakker,AB96}. In the following we are interested in
equilibrium or only small nonequilibrium, where the chemical potential
difference
$\mu_L-\mu_R\equiv U\ll
\Delta\sim k_BT_c$. In this case the  current has the form
\begin{equation}
I=\frac{2}{\hbar}\sum_i\sum_{\delta}
\frac{d\epsilon_{i\delta }}{d\phi}f_{i\delta }.
\label{e.cursc}\end{equation}
Here $f_{i\delta }$ is the occupation of the state. In
equilibrium it reduces to the Fermi distribution
$f_{i\delta }=[\exp(\epsilon_{i\delta }/k_B T)+1]^{-1}$. The index $i$
indicates different channels, which correspond to different directions
and locations of the bound state (Fig.\ \ref{jostrajboundsc}). In a
point contact of area $A$, there are $M=k_F^2A/4\pi$ channels, where
$k_F$ is the Fermi wave vector. In the normal state the current
is $I=MU/\pi\hbar$ \cite{Datta}.

Kulik and Omel'yanchuk derived in 1975 the following formula for
equilibrium current trough a superconducting point contact
\cite{KO} 
\begin{equation}
I=\frac{M\Delta}{\hbar}\sin\frac{\phi}{2}
\tanh\frac{\Delta\cos\frac{\phi}{2}}{2k_BT}.
\label{e.ko}\end{equation}
Their original derivation does not give much clue to understand the
result. Now this result can be straightforwardly understood based on the
formulas given above: all channels have the same bound state energies
$\epsilon=\pm\Delta\cos\frac{\phi}{2}$ and using the current
formula (\ref{e.cursc}) and the Fermi distribution  gives the
result (\ref{e.ko}).

As the temperature increases towards the superfluid transition
temperature $T_c$, the  Kulik-Omel'yanchuk result (\ref{e.ko}) reduces
to sinusoidal form 
\begin{equation}
I=I_c\sin\phi
\label{e.sinsoid}\end{equation}
with critical current $I_c=M\Delta^2/4\hbar k_BT_c$.

\section{$^3$He weak links}

\begin{figure}\label{diffusedos}
  \includegraphics[width=0.7\linewidth]{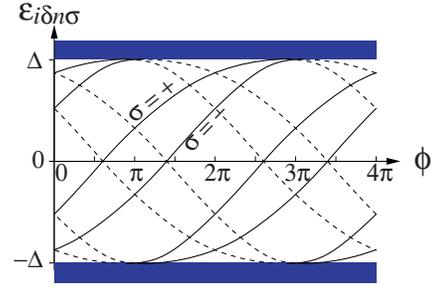}
  \caption{Example of bound states energies
$\epsilon_{i\delta n\sigma}(\phi)$ in a $^3$He weak link. Compared to
Figure \ref{diffusedossc}, the bound state energies have smaller
slopes and they are spin split ($\sigma=\pm 1$). 
Figure reprinted from Ref.\ \cite{dynlett}.
}
\end{figure}

Most of the discussion of superconducting weak links applies also to
weak links in $^3$He. There are a few differences.  In $^3$He the
Cooper pairs are in p-wave states, instead of the s-wave state
considered above. Any scattering breaks these pairs. Thus the
superfluid state is always suppressed near walls. This suppression also
affects the bound states energies in a weak link, as depicted in  Fig.\
\ref{diffusedos}.
Compared to the superconducting case, the slopes $d\epsilon/d\phi$ are
smaller in magnitude. Another difference is that there is spin
splitting of the energy states ($\sigma=\pm 1$).
Whereas the states in the superconducting case were doubly degenerate,
this degeneracy is generally broken in $^3$He. 

Because of the spin splitting the current formula (\ref{e.cursc}) has to
be  generalized to the form 
 \begin{equation}
I=\frac{1}{\hbar}\sum_i\sum_{\delta}\sum_{n}\sum_\sigma
\frac{d\epsilon_{i\delta n\sigma}}{d\phi}f_{i\delta n\sigma},
\label{e.curhelium}\end{equation}
where a factor of 2 is replaced by spin summation.
There is also additional summation over index $n$, which takes into
account that several bound states can occur at a given $\phi$.

The spin splitting can lead to current--phase relationships that differ
essentially from the standard sinusoidal form (\ref{e.sinsoid})
\cite{Yip}. Consider, for example, the case where
the spin split states are shifted relative to each other by 
a phase difference 
$\approx\pi$. It follows that the
leading sinusoidal components of the current from the two spin states
[$\propto\sin(\phi+\psi_\sigma)$] cancel each other. What is then left
are higher harmonics [$\propto\sin(n\phi+\psi_\sigma)$, $n=3,\ldots$].
Unusual current--phase relations are confirmed by 
calculations below.

The Cooper pairs of superfluid $^3$He have p-wave form \cite{Leggett}.
There are three orthogonal p-wave states: $p_x$, $p_y$, and $p_z$. The
spin state is triplet. There are three triplet states, which
conventionally are chosen as 
$-\uparrow\uparrow+\downarrow\downarrow$, 
$i\uparrow\uparrow+i\downarrow\downarrow$, and
$-\uparrow\downarrow+\downarrow\uparrow$.
Here we concentrate on the B-phase, where $p_x$ pairs have the first
spin state, $p_y$ pairs the second and $p_z$ pairs the third. However,
the spin coordinate axes ($x',y',z'$) are rotated  relative to the
orbital axes ($x,y,z$). The rotation angle is fixed to $104^\circ$,
but the axis $\hat{\bf n}$ of this rotation can vary.

In hydrodynamics of the B phase the mass ($\phi$) and
spin ($\hat{\bf n}$) degrees of freedom are independent.
The weak link acts as a nonlinear element that couples mass and spin,
as will be demonstrated below.

\subsection{Isotextural Theory}

\begin{figure}\label{PinholeCP3b}
  \includegraphics[width=0.99\linewidth]{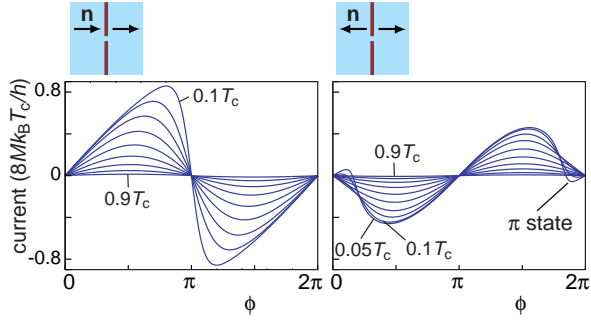}
  \caption{Current--phase relationships calculated for a pinhole. The
left hand panel is for parallel $\hat{\bf n}$ vectors on the two
sides of the junction. The right hand panel is for antiparallel
$\hat{\bf n}$'s that are perpendicular to the wall. At high temperature
the curves are sinusoidal (\ref{e.sinsoid}). At lower temperature the
curves become skew in the parallel case resembling the 
Kulik-Omel'yanchuk result (\ref{e.ko}).  The antiparallel case has
negative critical current $I_c$, or equivalently, is shifted by a phase
difference $\pi$. At very low temperature it develops an additional
kink that is known as
$\pi$ state. Figure adapted from Ref.\
\cite{Viljas2}. }
\end{figure}

Some current--phase relations calculated for a $^3$He-B pinhole are
shown in Fig.\ \ref{PinholeCP3b}.
Two cases are shown. First, the spin rotation axes $\hat{\bf n}$ on both
sided of the junction are the same, and second, the rotation axes have
opposite directions perpendicular to the wall of the pinhole.
The former case resembles the Kulik-Omel'yanchuk result (\ref{e.ko}):
at high temperatures $I(\phi)$ is sinusoidal but at lower temperatures
it becomes skew towards $\phi=\pi$. The case with antiparallel
$\hat{\bf n}$'s is also sinusoidal at high temperatures, but has
negative $I_c$. (Such a case is called a {\em $\pi$ junction} in
superconductivity.)  At temperatures around
$0.1 T_c$ it develops to a {\em $\pi$ state}. This means that the slope
of
$I(\phi)$ is positive at both 0 and $\pi$. 

\begin{figure}\label{BerkeleyPi}
  \includegraphics[width=0.99\linewidth]{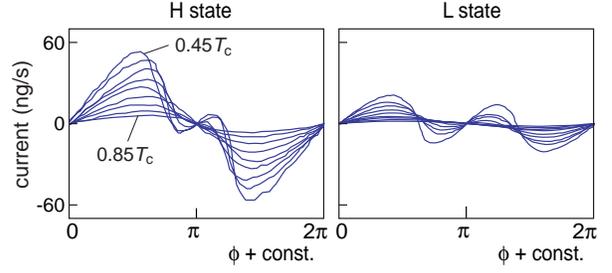}
  \caption{Measured current--phase relationships for a $65\times 65$
  array of 100 nm diameter apertures in 50 nm thick wall
  \cite{bistability}. In cooling
  through $T_{\rm c}$, the system randomly freezes to either H or L
  state. Only the relative value of the phase is determined
  experimentally. Figure adapted from Ref.\
\cite{bistability}. }
\end{figure}

Prior theoretical calculations, the $\pi$ state had been observed
experimentally. The experimental results of Ref.\  
\cite{bistability} are shown in Fig.\ \ref{BerkeleyPi}.
The experimental $I(\phi)$ is sinusoidal at high temperature but
develops a $\pi$ state at low temperature. Moreover, 
two different metastable states with different sets of $I(\phi,T)$
curves was observed. These are called H and L, corresponding to high and
low critical current. The two states appeared randomly in
cooling through
$T_c$, but the state selected remained stable during the stay in
the superfluid state. 

Let us consider a weak link under a constant chemical potential
difference $U=\mu_L-\mu_R$. According to the Josephson relation
\begin{equation}
\frac{d\phi}{dt}=\frac{2U}{\hbar}
\label{e.jos2}\end{equation}
the phase $\phi$ increases linearly in time. This implies that the
bound state energies (Fig.\ \ref{diffusedos}) are continuously shifting
up or down. During shifting the occupations of these states can 
change in collisions with other quasiparticles. Since the collisions
are rare, this thermalizes the occupations only at small
$U<\hbar/\tau$, where $\tau$ is the scattering time. At larger
bias, the occupations of the bound states remain essentially constant
during the shift from $-\Delta$ to  $+\Delta$, or from $+\Delta$ to 
$-\Delta$. The occupations of these states are then fixed to the
thermal equilibrium occupations at the starting energies, $-\Delta$ or 
$+\Delta$, respectively.

\begin{figure}\label{IPfitS_8ed}
   \includegraphics[width=0.65\linewidth]{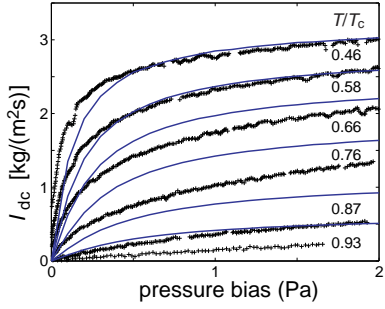}
   \caption{Average current vs. pressure bias. The data points are
   experimental results from Ref.\
   \cite{Steinhauer}, and the lines are theoretical results from
   Ref.\ \cite{Viljasoma}. The pressure bias of 
   1 Pa corresponds to $U=4.7\times10^{-3}k_BT_c$. 
   Figure reprinted from Ref.\
   \cite{Viljasoma}.}
\end{figure}

It is now possible to calculate the current using Eq.\
(\ref{e.curhelium}).  The
resulting time-averaged current $I_{\rm dc}$ is plotted  in 
Fig.\ \ref{IPfitS_8ed}.
At large bias ($U\gg\hbar/\tau$) the occupation are
determined by gap edges and the current therefore becomes
independent of the bias $U$. At small bias ($U\ll\hbar/\tau$) the
scattering has time to preserve a nearly thermal distribution, and
the deviation from equilibrium distribution as well as the current is
linear in
$U$. Note that the equilibrium current (Fig.\ \ref{PinholeCP3b}) is
oscillating, and therefore does not contribute to the time-averaged
current.  

Fig.\ \ref{IPfitS_8ed}.
shows also experimental data by Steinhauer et al. \cite{Steinhauer}.
The theory can reasonably be fitted to this data using the scattering
time as the only fitting parameter. The agreement is surprisingly good
taking into account that the single aperture in the experiment
(dimensions $7.8\ \mu{\rm m}\times 0.27\ \mu{\rm m}$) is large compared
to the superfluid coherence length $\xi_0=77$ nm, whereas the pinhole
theory makes just the opposite assumption. 

The current at finite bias was also measured for the same array of 
apertures as the equilibrium current in Fig.\ \ref{BerkeleyPi}. 
\begin{figure}\label{HLstatedyn}
   \includegraphics[width=0.99\linewidth]{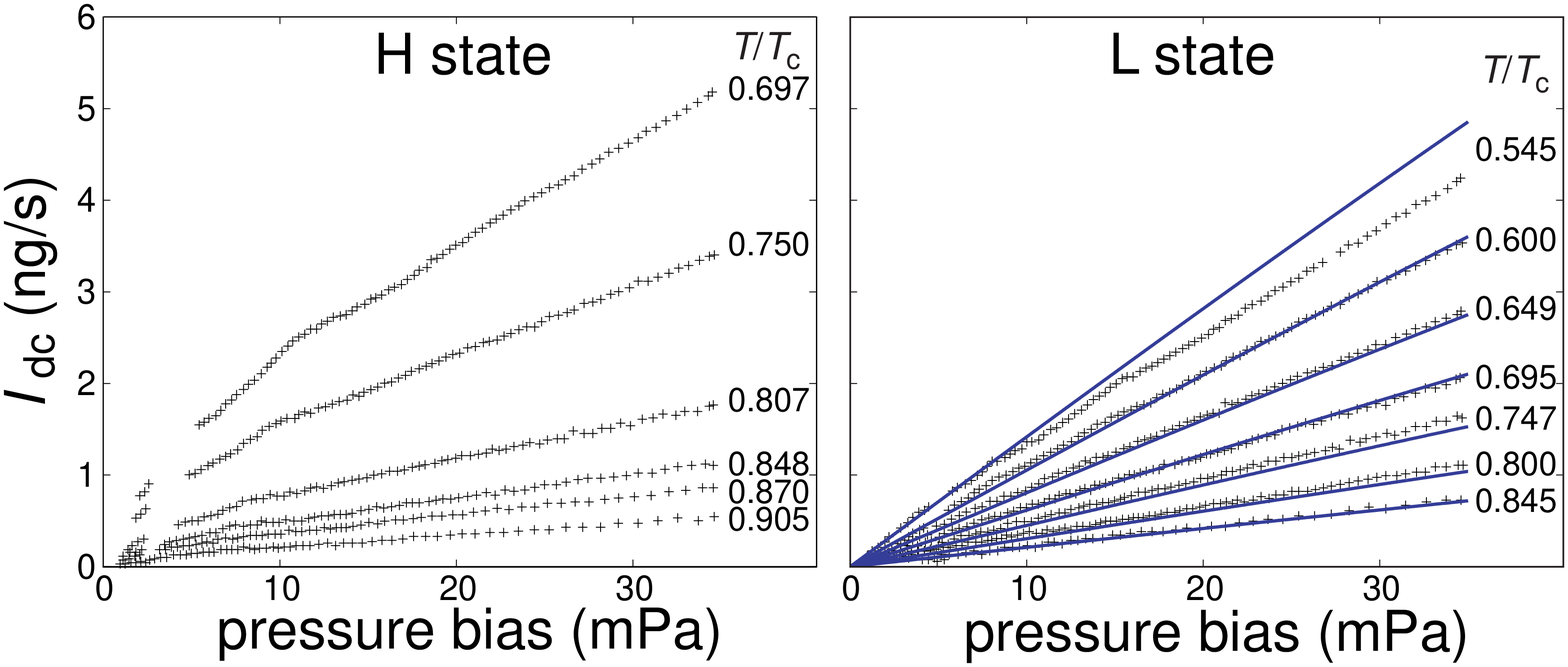}
   \caption{Average current vs. pressure at low pressure
   bias. The data points are experimental results from Ref.\
   \cite{Simmonds}
  for the same $65\times 65$
  array of  apertures as the equilibrium currents in
   Fig.\ \ref{BerkeleyPi}. The lines drawn on the right hand side panel
   are the results of isotextural theory, and fitted with a single
   parameter (the relaxation time) as in Fig.\ \ref{IPfitS_8ed}.
  According to isotextural theory, the same lines should fit also the H
  state, which clearly is not the case.
   Figure adapted from Ref.\
   \cite{dynlett}.}
\end{figure}
The results are shown in  
Fig.\ \ref{HLstatedyn}. The current is different for the H and L
states. The current in the L state is in reasonable agreement with
theory  using again the relaxation time as the only adjustable
parameter.

As a summary thus far, we can say that theory and experiment are in
reasonable agreement. Both show bistability (H and L state), $\pi$
states, and similar dissipative currents $I_{\rm dc}(U)$. The
temperature dependencies agree as well the order of magnitudes of all
quantities. 

In spite of this success, closer examination reveals some problems in
the theory. 1) Only one of theoretical bistable states shows $\pi$
state  (Fig.\ \ref{PinholeCP3b}). More detailed analysis reveals that
this problem cannot be removed by considering other configurations of 
$\hat{\bf n}$'s \cite{Viljas2}.
2) The theoretical $\pi$ state occurs at a much lower temperature and is
much weaker than in experiments (compare  Figs.\
\ref{PinholeCP3b} and \ref{BerkeleyPi}).
3) The theory does not predict any difference in $I_{\rm dc}(U)$ between
the H and L states. This is because different $\hat{\bf n}$
configurations mainly shift the bound states energies
$\epsilon_{i\delta n\sigma}$ along the $\phi$ axis but do
not much affect the shape of the $\epsilon_{i\delta n\sigma}(\phi)$
curves. In running $\phi$ this only affects the instantaneous current
$I(t)$, but the average 
$I_{\rm dc}$ is unchanged. 

\subsection{Anisotextural Theory}

All the problems above can be explained by a single new concept.
If the Josephson coupling is strong, it can change the spin-part of
the order parameter on both sides. We call this {\em anisotextural}
effect, since the $\hat{\bf n}$ orientations are commonly known as
texture. 

As an example consider the case of parallel $\hat{\bf n}$'s. Then there
is no spin splitting and the bound state at phase difference 
$\phi=\pi$ lies at zero energy, $\epsilon=0$. The thermal occupation
of this doubly degenerate state is $f=1/2$. If we now allow spin
splitting, the $\epsilon=0$ state can split into two
with positive and negative energies (Fig.\ \ref{diffusedos}). Taking
into account that the occupation of the negative energy state is
larger than that of the positive energy state, this leads to a
reduction of energy. Thus the spin splitting can take place
spontaneously.

\begin{figure}\label{JosIdeaPuolikas2}
  \includegraphics[width=0.99\linewidth]{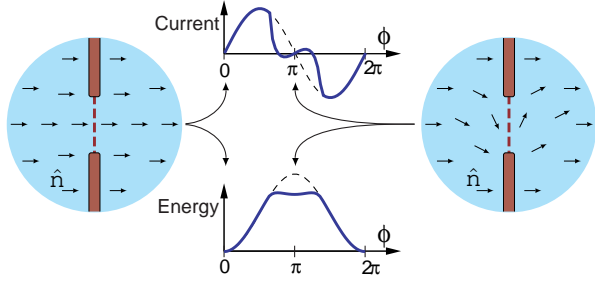}
  \caption{Principle of the anisotextural Josephson effect.
  The panels on the left and right hand sides depict an array of
  apertures
  and the configurations of the spin-rotation axis $\hat{\bf n}$.
  Assuming there is no spin
  structure ($\hat{\bf n}={\rm constant}$) at the phase difference
  $\phi=0$,
  the $\hat{\bf n}$ texture changes spontaneously when $\phi$ increases
  to $\pi$. This leads to a reduction of energy and to a positive slope
of
   $I(\phi)$ at $\phi=\pi$, as shown by the change from dashed lines to
  solid lines in the current and energy plots.}
\end{figure}

The anisotextural effect as a function of phase is
shown in Fig.\ \ref{JosIdeaPuolikas2}.
Starting from the completely spin-symmetric situation at $\phi=0$, the 
$\hat{\bf n}$ texture changes spontaneously as $\phi$ increases
towards $\pi$. The reduction of the Josephson coupling energy 
$F_{\rm J}$ is associated with a change in $I(\phi)$
since 
\begin{equation} \label{e.enederiv}
I=\frac{2}{\hbar}\frac{\partial F_{\rm J}}{\partial\phi}.
\end{equation}
Thus a $\pi$ state develops if the reduction of $F_{\rm J}$ is
sufficient to produces a local minimum of $F_{\rm J}(\phi)$ at 
$\phi=\pi$.  
 
\begin{figure}\label{CPait}
  \includegraphics[width=0.99\linewidth]{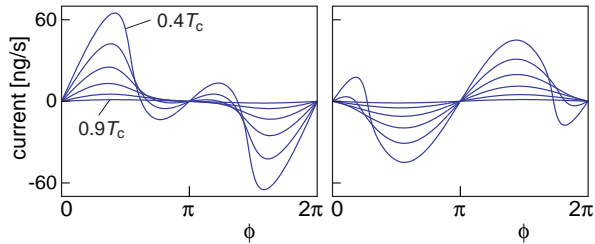}
  \caption{Current--phase relationships calculated using
  the anisotextural
  model (\ref{e.fait}). The theory contains one free parameter
  $\eta_\infty$, and
  the result should be compared with the experimental one in 
  Fig.\ \ref{BerkeleyPi}.  Figure adapted from Ref.\ \cite{Viljas2}.}
\end{figure}

An additional contribution to energy arises from the bending of
$\hat{\bf n}$ (right hand panel of Fig.\
\ref{JosIdeaPuolikas2}). In a simple model the total energy is written
as  
\begin{equation}
\label{e.energy}
F[\eta] = F_J(\eta_0,\phi)
+\textstyle{\frac{1}{2}}K\int d^3r \vert\nabla\eta\vert^2.
\label{e.fait}\end{equation}
The first term is the Josephson coupling energy that depends on
$\phi$ and on the tilting angle $\eta_0$ of $\hat{\bf n}$ at the weak
link. The second term is a gradient energy for the tilting angle in the
bulk. In comparison to experiments, the only free parameter is the
tilting angle
$\eta_\infty$ on one side far away from the junction, which is
difficult to calculate because of the complicated shape of the
experimental cell. Using the freedom to adjust $\eta_\infty$, it is
possible to generate the current--phase relations shown in Fig.\
\ref{CPait}.
This should be compared with the experimental curves in Fig.\
\ref{BerkeleyPi}.

Next we consider the anisotextural effect at
constant bias $U$. Finite $U$ means that the phase changes at constant
rate  [Eq.\ (\ref{e.jos2})]. As a result of the anisotextural effect,
the
$\hat{\bf n}$ texture oscillates at the angular frequency
$\omega=2U/\hbar$. This oscillation of the spin structure generates
spin waves, which radiate out of the junction. 

In order to make an quantitative theory, we need to add one more
contribution to the energy (\ref{e.fait}) so that the total energy is
\begin{eqnarray}
F[\eta,S] = F_J(\eta_0,\phi)
+\textstyle{\frac{1}{2}}K\int d^3r \vert\nabla\eta\vert^2\nonumber\\
+\textstyle{\frac{\gamma^2}{2\chi}}\int d^3r S^2.
\end{eqnarray}
The last therm is the energy associated with net spin density $S$,
where 
$\gamma$ is the gyromagnetic ratio and $\chi$ the magnetic
susceptibility \cite{Leggett2}. Here $S$ and $\eta$ are conjugate
variables. Writing  the Hamiltonian equations, this leads to a wave
equation describing the spin waves and to a boundary condition for
them. It turns out to be a quite standard radiation problem. The
radiated power has the frequency dependence
\begin{equation}
P_{\rm rad}\propto \frac{\omega^2}{1+(r_0\omega/c)^2},
\end{equation}
where $r_0$ is the radius of the Josephson array, and $c$ the spin
wave velocity. This expresses that there is little radiation for
wave lengths longer than the source, but more at shorter wave
lengths.  

The dissipated power now has to come from the dc current, $P = U
I_{\rm dc}$. Thus there is additional dc current due to the spin wave
radiation. Including also the prefactors we get for it the
expression
\begin{equation}
I_{\rm dc, rad}=\frac{P_{\rm rad}}{U}=\frac{2}{\hbar}
\frac{[J_{\rm sp}(\eta_\infty)]^2}{4\pi cK}
\frac{\omega }{1+(\omega r_0/c)^2}.
\end{equation}
Here the equilibrium spin current $J_{\rm sp}(\eta_\infty)$ plays a role
of a coupling constant between mass and spin variables. Most
interestingly, this coupling is different for the theoretical H and L
states. Putting in numbers we see that the theory explains
approximately one third of the observed H-L difference shown in Fig.\
\ref{BerkeleyPi}. (The comparison is shown in Ref.\ \cite{dynlett}.)
This is not completely satisfactory, but we must remember that all
parameters of the theory were fixed before the comparison.

Models to explain the observed $I_{\rm dc}(U)$ have also been
suggested in Ref.\ \cite{Simmonds}. These models differ essentially
from the present ones. In particular, an A-phase-like distortion of 
the order parameter is associated with the linear
part and a quasiparticle radiation mechanism with the nonlinear part of
$I_{\rm dc}(U)$.

\section{Conclusion}

We have seen that coupling between mass and spin variables is essential
for understanding the Josephson phenomena in superfluid $^3$He.
With spin splitting of bound states together with the anisotextural
effect it is possible to explain most of the experimental results of
both
$I(\phi)$ and $I_{\rm dc}(U)$. The theory presented above uses the
pinhole approximation. For more accurate results it is necessary
to consider larger apertures. This would require selfconsistent
determination of the order parameter within the aperture, which is
calculationally demanding and has been done only in limiting cases
\cite{TKS,glviljas}.
We expect that this might lead to qualitatively same type of results
as the anisotextural effect for pinhole arrays, in particular, to
enhanced
$\pi$ states and to a texture dependence of $I_{\rm dc}(U)$. This might
explain the remaining differences between theory and experiment, and the
experimental results in single apertures \cite{singleaperture}.

The anisotextural phenomena depend strongly on many parameters such as
array size, geometry of the experimental cell and magnetic field.
We hope that these could be tested in future experiments.

\end{document}